# A Step towards an Easy Interconversion of Various Number Systems

Shahid Latif, Rahat Ullah, Hamid Jan

*Department of Computer Science and Information Technology*
**Sarhad University of Science and Information Technology (SUIT)**
**Peshawar 25000 Pakistan**
Shahid.csit@suit.edu.pk
rahat.csit@suit.edu.pk
hod.csit@suit.edu.pk

ABSTRACT
Any system that is used for naming or representing numbers is a number system, also known as numeral system. The modern civilization is familiar with decimal number system using ten digits. However digital devices and computers use binary number system instead of decimal number system, using only two digits namely, 0 and 1 based on the fundamental concept of the decimal number system. Various other number systems also used this fundamental concept of decimal number system, for example octal system and hexadecimal number systems using eight and sixteen digits respectively. The knowledge of number systems and their inter conversion is essential for understanding of computers. More over, successful programming for digital devices requires a precise understanding of data formats, number systems and their inter conversion. The inter conversion (a process in which things are each converted into the other) of number system requires allot of time and techniques to expertise. In this paper the interconversion of four most common number systems is taken under the consideration in tabulated form. It is a step towards the easy interconversion of theses number systems to understand as well as memorise it. The four number systems are binary, octal, decimal and hexadecimal.

## 1. INTRODUCTION

In the digital world especially computer and information technology, normally we requires a working knowledge of various number systems, four of which are the most common such as binary, octal, decimal and hexadecimal. More specifically, the use of the microprocessor requires a working knowledge of binary, decimal and hexadecimal numbering system [1]. Computers communicate and operate in binary digits 0 and 1; on the other hand human beings generally use the decimal systems with ten digits 0-9. Other number systems are also used in digital systems, such as octal with eight digits i.e. 0-7 and hexadecimal system with digits from 0-15. In hexadecimal system, digits 10-15 are designated as A through F. Respectively to avoid confusion with the decimal numbers, 10 to 15 [2].

In data communication a simple signal by itself does not carry information any more than a straight lines conveys words. The signal must be manipulated so that it contains identifiable changes that are recognizable to the sender and receiver as representing the information intended. First the information must be translates into agreed-upon patterns of 0s and 1s, for example, using ASCII. Also, data stored in the computer are in the form of 0s and 1s. To be carried from one place to another, data are usually converted to digital signals. Some times we need to convert an analog signal (such as voice in a telephone conversation) into a digital signal and vice versa [3].

So, in many applications we deal with the ineterconversion of number systems. There are various techniques that used for these inter conversions.

Remember, all number systems are interconvertable. But each conversion i.e. from one number system to another often takes place in a different way, using different techniques. So it becomes very tedious for beginners to overcome this difficulty and understand these conversions in short time.

In this particular paper, we introduce a tabulated format for these conversions, which covers all these interconversions in only three steps, taking approximately one contact hour of the lecture. While, in earlier approaches we have to use more than 20 steps to perform all these conversions.

This paper is organized in such a way that it consist of five sections. Section one covers the brief introduction of the number systems, need of number systems and their interconversion and easy approach to it. Section two is the overview of the number systems and their representations. Section three describes all the conversion techniques (both for integral and fraction part of the numbers) frequently used so far. Section four contains and the proposed tabulated form for interconversion processes, while last one section conclude the paper.

## 2. OVERVIEW OF NUMBER SYSTEMS

When humans are speaking to one another, they speak in a particular language. This language is made of words and letters. Although we type words and letters in the computer, the computer does not understand the words and letters. Rather, those words and letters are translated into numbers. Computers "talk" and understand in numbers. Although many students know the decimal (base 10) system, and are very comfortable with performing operations using this system, it is important for students to understand that the decimal system is not the only system. By studying other number systems such as binary (base 2), octal (base 8), and hexadecimal (base 16), students will gain a better understanding of how number systems work in general.

### 2.1 Digits

Before numbers are converted from one number system to another, the digit of a number system must be understood. The first digit in any numbering system is always a zero. For example, a base 2 (binary) numbers contains 2 digits: 0 and 1, a base 8 (octal) numbers contains 8 digits: 0 through 7 and so on. Note that a base 10 (decimal) numbers does not contain 10 digits, just as base 8 numbers does not contain an 8 digit.

Once the digits of a number system are understood, larger numbers are constructed by using positional notation. As in decimal the position to the left of the units position was the tens position, the position to the left of the tens position was the hundreds position and so forth. Here, the units position has a weight of $10^0$, or 1; the tens position has a weight of $10^1$, or 10; and the hundreds position has a weight of $10^2$, or 100. The exponential powers of the positions are critical for understanding numbers in other numbering systems. The position to the left of the radix point is always the unit's position in any number system. For example the position to the left of the binary point is always $2^0$, or 1; the position to the left of the octal point is always $8^0$, or 1 and so on.

The position to the left of the units position is always the number base raised to the first power; i.e. $2^1$, $8^1$ and so on.

## 2.2 Number representation

A number in any base system can be represented in a generalized format as follows:

$N = A_n B^n + A_{n-1} B^{n-1} + - - - + A_1 B^1 + A_0 B^0$, where
N = Number, B=Base, A= any digit in that base

For example number 154 can be represented in various number systems as follows:

| Decimal | 154 | $1\times 10^2 + 5\times 10^1 + 4\times 10^0$ = 100 + 50 + 4 | 154 |
|---|---|---|---|
| Binary | 10011010 | $1\times 2^7 + 0\times 2^6 +...+ 0\times 2^0$ = 128+0+0+16+8+0+2+0 | 154 |
| Octal | 232 | $2\times 8^2 + 3\times 8^1 + 2\times 8^0$ = 128 + 24 + 2 | 154 |
| Hexa-decimal | 9A | $9\times 16^1 + A\times 16^0$ = 144 + 10 | 154 |

## 2.3 Most Significant Digit and Least Significant Digit

The MSD in a number is the digit that has the greatest effect on that number, while The LSD in a number is the digit that has the least effect on that number.
Look at the following examples:

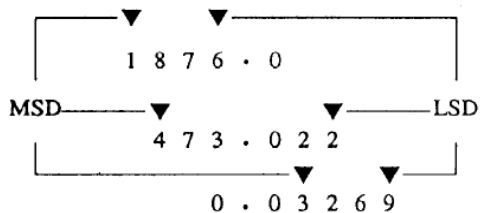

You can easily see that a change in the MSD will increase or decrease the value of the number in the greatest amount, while changes in the LSD will have the smallest effect on the value.

## 2.4 Decimal Number System

The decimal number system is known as international system of numbers [14]. It is also called base ten or occasionally denary number system. It has ten as its base. It is the numerical base most widely used by modern civilization [7].
Decimal notation often refers to a base-10 positional notation; however, it can also be used more generally to refer to non-positional systems. Positional decimal systems include a zero and use symbols (called digits) for the ten values (0, 1, 2, 3, 4, 5, 6, 7, 8, and 9) to represent any number, no matter how large or how small.

Let's examine the decimal (base 10) value of 427.5. You know that this value is four hundred twenty-seven and one-half. Now examine the position of each number:

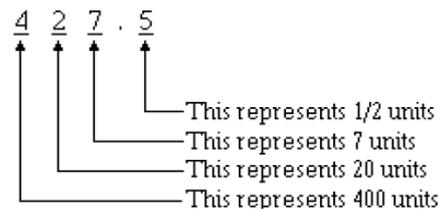

Each digit has its own value (weight) as described in the above figure. Now let's look at the value of the base 10 number 427.5 with the positional notation line graph:

| | Radix Point | | | |
|---|---|---|---|---|
| $10^2$ | $10^1$ | $10^0$ | . | $10^{-1}$ |
| 4 | 2 | 7 | . | 5 |

$10^2$ = 4 × 100, or 400
$10^1$ = 2 × 10, or 20
$10^0$ = 7 × 1, or 7
$10^{-1}$ = 5 × .1, or .5

You can see that the power of the base is multiplied by the number in that position to determine the value for that position. All numbers to the left of the decimal point are whole numbers or integers, and all numbers to the right of the decimal point are fractional numbers.

## 2.5 Binary Number System

The number system with base (or radix) 2, is known as the binary number system. Only two symbols are used to represent numbers in this system and these are 0 and 1, these are known as bits. It is a positional system i.e. every position is assigned a specific weight. Moreover, it has two parts the Integer and fractional, set a part by a radix point. For example $(1101.101)_2$
In binary number system the left–most bit is known as most significant bit (MSB) and the right–most bit is known as the least significant bit (LSB), similar to decimal number system. The following graph shows the position and the power of the base:

$$2^3 \; 2^2 \; 2^1 \; 2^0 \, . \, 2^{-1} \; 2^{-2} \; 2^{-3}$$

The arithmetic operations such as addition, subtraction, multiplication and division of decimal numbers can be performed on binary numbers. Also binary arithmetic is much simpler than decimal arithmetic because here only two digits, 0 and 1 are involved.

## 2.6 Octal Number System

As its name reveal (octal = 8), the number system with base eight (8) is known as the octal number system. In this system eight symbols, 0, 1, 2, 3,4,5,6, and 7 are used to represent the number. Similar to decimal and binary number systems, it is also a positional system; the octal system uses power of 8 to determine the value of a number's position. The following graph shows the positions and the power of the base:

$$8^3 \; 8^2 \; 8^1 \; 8^0 \, . \, 8^{-1} \; 8^{-2} \; 8^{-3}$$

Octal number has two parts: Integer and fractional, set a part (separated) by a radix point, for example $(6327. 4051)_8$
It is highly tedious to handle long strings of binary numbers while entering into the digital systems. It may cause errors also. Therefore,

octal numbers are used for entering binary data and displaying certain information in short.

## 2.7 Hexadecimal Numbering System

Hexadecimal number system is very popular in computer uses. The base for hexadecimal number system is 16 which require 16 distinct symbols to represent the number. These are numerals 0 to 9 and alphabets A to F. This is an alphanumeric number system because its uses both alphabets and numerical to represent a hexadecimal number. Hexadecimal numbers are 0,1,2,3,4,5,6,7,8, 9,A,B,C,D,E,F. Any number in hexadecimal number system can be represented as (B52.AC3)16.

It has also two parts i.e. integer and fractional. Like the binary, octal, and decimal systems, the hexadecimal number system is a positional system. Powers of 16 are used for the positional values of a number.

The following graph shows the positions:

$$16^3 \; 16^2 \; 16^1 \; 16^0 \; . \; 16^{-1} \; 16^{-2} \; 16^{-3}$$

The most significant and least significant digits will be determined in the same manner as the other number systems.

## 3 CONVERSION BETWEEN NUMBER SYSTEMS

The most common number systems are the decimal, binary, octal and hexadecimal, we have discussed so far. Now we have to check that how any number can be converted from one to another number system. Number systems are udergiven in the ascending order as,

Binary
Octal
Decimal
Hexadecimal

A given number in any of the above number systems may consist of two parts i.e. the integer part and the fraction part. Each part some times, required a different technique for conversion. In other words, in case of fractions (fraction numbers) the conversion process requires additional techniques. So as a whole, more than 20 various techniques are used for these inter conversions, which are enlisted below.

**Integral part of a numbers**

Binary – to – octal
Octal – to – binary
Binary – to – decimal
Decimal – o – binary
Binary- to – hexadecimal
Hexadecimal – to – binary
Octal – to – decimal
Decimal –to – octal
Octal –to – hexadecimal
Hexadecimal –to- octal
Decimal – to – hexadecimal
Hexadecimal –to – decimal

**Fractions**

Binary fraction – to – octal
Octal fraction– to – binary
Binary fraction – to – decimal
Decimal fraction – to – binary
Binary fraction - to – hexadecimal
Hexadecimal fraction – to – binary
Octal fraction – to – decimal
Decimal fraction –to – octal
Octal fraction –to- hexadecimal
Hexadecimal fraction –to- octal
Decimal fraction – to – hexadecimal
Hexadecimal fraction –to – decimal

It is obvious, that the beginners will be very disharted/ disappointed to use such a lot of techniques for the conversion in short time of one contact hour or so.

## 4 TABULATED FORMATE

As we know that the decimal number system is the most common of the above mentioned number systems, because it is widely used in mathematics and our daily life calculations based on this number system, so we start the conversion from the decimal number system to the remaining systems. Here a word "other" is used for those number systems which are other than the mentioned one.

We can perform the conversion between different number systems in three steps,

Step: 1    A) From Decimal number system → to → other number systems [Binary, Octal, Hexadecimal]
          B) From Other number systems [Binary, Octal, Hexadecimal] → to → Decimal number system

In step: 1 all the conversion processes related to decimal number system are covered. So we will not use the conversion from/to decimal number system to/from others, anymore.

Step: 2    A) From Binary number system → to → other [Octal, Hexadecimal] number systems
          B) From Other [Octal, Hexadecimal] number systems → to → Binary number system

In step: 2 all the conversion processes related to binary number system are covered. So we will not use the conversion from/to binary number system to/from others, anymore.

Step: 3    A) From octal number system → to → hexadecimal number systems
          B) From hexadecimal number systems → to → octal number system

**Step: 1**

A) Conversion from Decimal number system → to → other number systems [Binary, Octal, Hexadecimal]

To convert a given number from decimal number system to any other number system, follow these steps:
1. Divide the decimal number by r i.e. base of the other system (2, 8, or 16). Remember the quotient and the remainder of this division.
2. After that, divide the quotient (from the first division) by r, again remembering the quotient and the remainder.
3. Keep dividing your new quotient by r until you get a quotient of 0. After each division, keep track of the remainder.
4. When you reach a quotient of 0, the remainders of all the divisions (written in reverse order) will be the equivalent number in base r number system.
[Reverse order mean that, the first remainder that you got in step-1 will be the least significant digit (LSD) of the number in base r number system].

B) Conversion from Other number systems [binary, octal, hexadecimal] → to → decimal number system

The conversion process from other number systems [i.e. binary, octal, and hexadecimal] to decimal number system has the same procedure.

Here any binary number can be converted into its equivalent decimal number using the weights assigned to each bit position. Incase of binary the weights are $2^0$ (Units), $2^1$ (twos), $2^2$ (fours), $2^3$ (eights), $2^4$ (sixteen) and so on. Similarly in case of octal the weights are $8^0$, $8^1$, $8^2$, $8^3$, $8^4$ and so on. For hexadecimal the weights are $16^0$, $16^1$, $16^2$, $16^3$, $16^4$ and so on. Here few steps are given which are helpful in faster and easy conversion of other systems to decimal number system.
1. Write the given (i.e. 2, 8, or 16) base number
2. Write the corresponding weight $x^0$, $x^1$, $x^2$, $x^3$,…, under each digit.
3. Cross out any weight under a 0 (means that any 0 involve in given number).
4. Add the remaining weights.

**Step: 2**

A) Conversion from Binary number system → to → other [Octal, Hexadecimal] number systems

A binary number can be converted to octal number and hexadecimal number by replacing method. The binary digits are grouped by threes (in case of converting to octal) and fours (in case of converting to hexadecimal) respectively by starting from the decimal point and proceeding to the left and to the right. Add leading 0s (or trailing zeros to the right of decimal point) to fill out the last group of three or four if necessary. Then replace group of three or group of four with the equivalent octal digit.

B) Conversion from Other [Octal, Hexadecimal] number systems → to → Binary number system

For some computers to accept octal or hexadecimal data, the octal or hexadecimal digits must be converted to binary. This process is the reverse of binary to octal and hexadecimal conversion. To convert a given (octal or hexadecimal) number to binary, write out the number and then write below each digit the corresponding three-digit binary-coded octal equivalent (in case of converting from octal) or four-digit binary-coded hexadecimal equivalent (in case of converting from hexadecimal). A fraction in both octal and hexadecimal is converted to binary in the same manner.
The below table shows the equivalent in other number system of first 16 decimal numbers:

Table: conversion table- Decimal, Hexadecimal, Octal, Binary

| Dec | Hex | Oct | Bin |
|---|---|---|---|
| 0 | 0 | 000 | 0000 |
| 1 | 1 | 001 | 0001 |
| 2 | 2 | 002 | 0010 |
| 3 | 3 | 003 | 0011 |
| 4 | 4 | 004 | 0100 |
| 5 | 5 | 005 | 0101 |
| 6 | 6 | 006 | 0110 |
| 7 | 7 | 007 | 0111 |
| 8 | 8 | 010 | 1000 |
| 9 | 9 | 011 | 1001 |
| 10 | A | 012 | 1010 |
| 11 | B | 013 | 1011 |
| 12 | C | 014 | 1100 |
| 13 | D | 015 | 1101 |
| 14 | E | 016 | 1110 |
| 15 | F | 017 | 1111 |

**Step: 3**

A) Conversion from Octal number system → to → Hexadecimal number systems

The conversion from octal number system to hexadecimal number system is a two-step procedure using binary as an intermediate base. Octal is converted to binary and then binary to hexadecimal, grouping the binary digits into groups of fours and add zeros where needed to complete groups, which correspond each to a hexadecimal digit.

B) Conversion from Hexadecimal number systems → to → Octal number system

The conversion from hexadecimal number system to octal number system is the reversal of the same algorithm as declared in first part of step-3. Reverse the previous algorithm to achieve the conversion.

**Table**

The following table reveals the conversion between the four number systems in three steps along the methods of conversion (for integers and fractions).
Remember, in step 2 and 3 conversion technique (method) for integer and fraction part is not mentioned. It is so, because in both steps, same method is used for integers and fractions conversion.

Table: conversion between Decimal, Binary, Octal and Hexadecimal along the conversion techniques

| Step No: | Part-A | Part-B |
|---|---|---|
| Step 1 | Decimal to others [binary, octal, hexadecimal] (==)10 →(==)2,8,16 **Integer:** repeated division method **Fraction:** repeated multiplication method | Others [binary, octal, hexadecimal] to decimal (==)2,8,16 →(==)10 **Integer:** sum of [(+ve weights)×(integer)] **Fraction:** sum of [(-ve weights) ×(fraction)] |
| Step 2 | binary to other [octal, hexadecimal] (==)2 →(==)8,16 **To octal:** replace group of 3-binary bits by octal digit **To hex:** replace group of 4-binary bits hexadecimal digit (same method for both integral and fraction part) | Other [octal, hexadecimal] to binary (==)8,16 → (==)2 **From octal:** replace each octal digit by 3-bit binary **From hex:** replace each hexadecimal digit by 4-bit binary (same method for both integral and fraction part) |
| Step 3 | octal to hexadecimal (==)8 →(==)16 Direct conversion not applicable Octal → Binary → Hexadecimal | hexadecimal to octal (==)16 → (==)8 Direct conversion not applicable Hexadecimal → Binary → Octal |

**CONCLUSION**

In this paper we propose an easy, short and simple approach (using a single table) to the complete interconversion of various numbers from the four most common number systems used in the digital world specially computer technology. Remember that, these four number systems are not the only number systems used in digital world, but are the very common and frequently used in most of the digital technologies and devices. The complete inter conversion takes a lot of time to understand and memorize all the techniques involved. From this paper we conclude that this is simply shorthand to the interconversion used in digital technology providing a rapid practice to the interconversion between various number systems and ease in memorizing all the techniques used for these conversions. As a future work, the conversion table proposed in this paper may be enhanced by using more number systems and their interconversions. Also newer conversion techniques can be added in it.